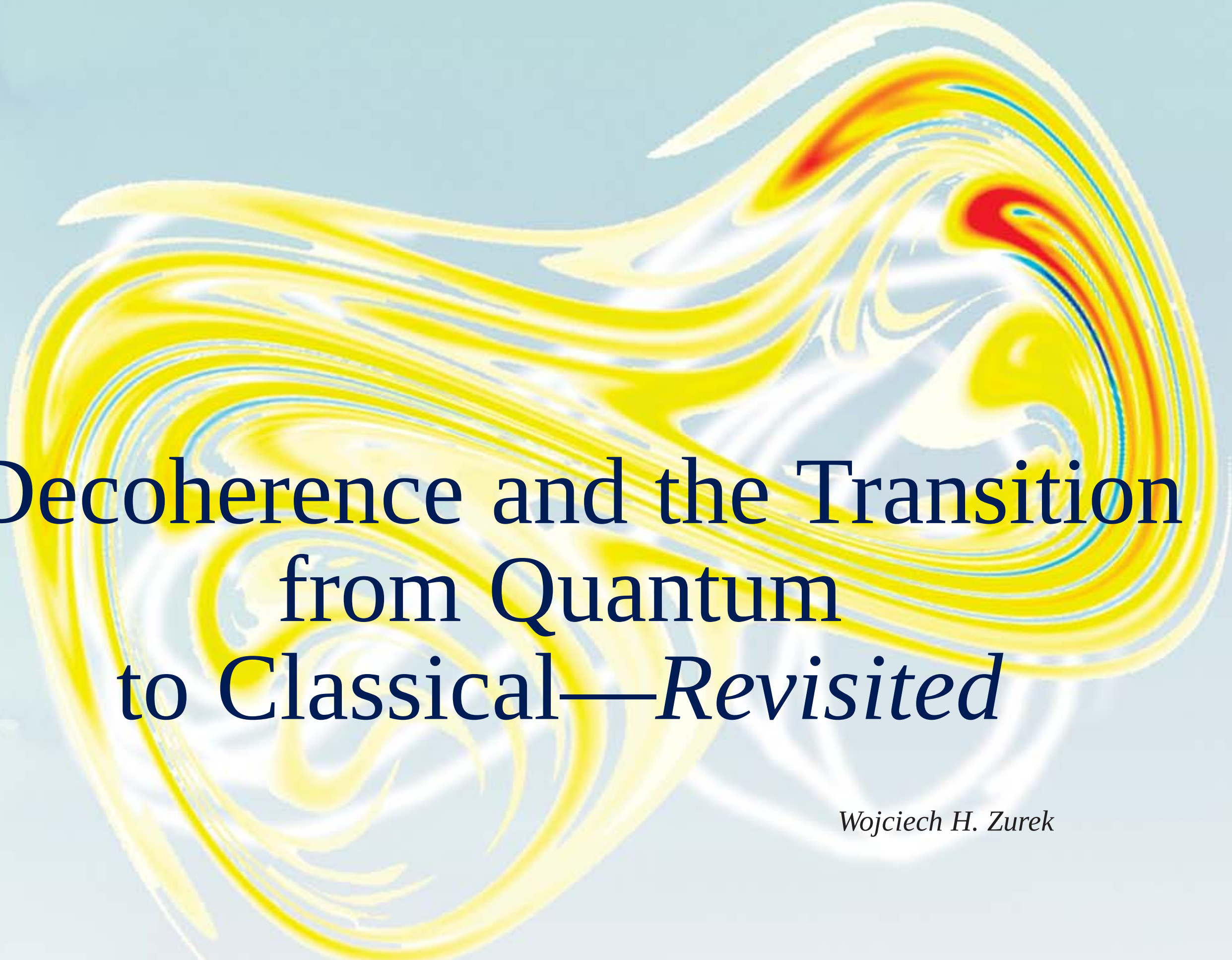

# Decoherence and the Transition from Quantum to Classical—*Revisited*

*Wojciech H. Zurek*

This paper has a somewhat unusual origin and, as a consequence, an unusual structure. It is built on the principle embraced by families who outgrow their dwellings and decide to add a few rooms to their existing structures instead of starting from scratch. These additions usually “show,” but the whole can still be quite pleasing to the eye, combining the old and the new in a functional way.

What follows is such a “remodeling” of the paper I wrote a dozen years ago for *Physics Today* (1991). The old text (with some modifications) is interwoven with the new text, but the additions are set off in boxes throughout this article and serve as a commentary on new developments as they relate to the original. The references appear together at the end.

In 1991, the study of decoherence was still a rather new subject, but already at that time, I had developed a feeling that most implications about the system’s “immersion” in the environment had been discovered in the preceding 10 years, so a review was in order. While writing it, I had, however, come to suspect that the small gaps in the landscape of the border territory between the quantum and the classical were actually not that small after all and that they presented excellent opportunities for further advances.

Indeed, I am surprised and gratified by how much the field has evolved over the last decade. The role of decoherence was recognized by a wide spectrum of practicing physicists as well as, beyond physics proper, by material scientists and philosophers.

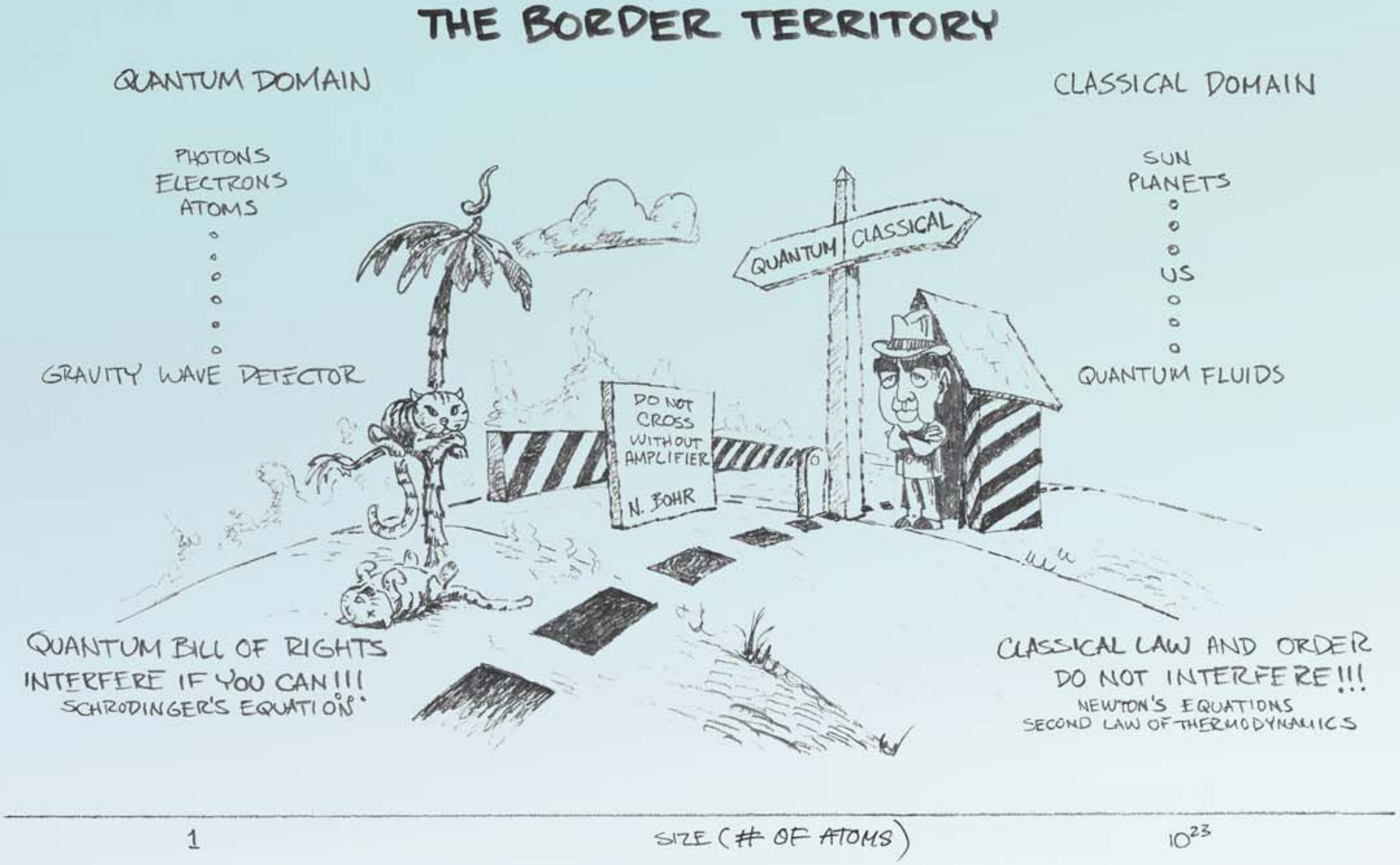

The study of the predictability sieve, investigations of the interface between chaotic dynamics and decoherence, and most recently, the tantalizing glimpses of the information-theoretic nature of the quantum have elucidated our understanding of the Universe. During this period, Los Alamos has grown into a leading center for the study of decoherence and related issues through the enthusiastic participation of a superb group of staff members, postdoctoral fellows, long-term visitors, and students, many of whom have become long-term collaborators. This group includes, in chronological order, Andy Albrecht, Juan Pablo Paz, Bill Wootters, Raymond Laflamme, Salman Habib, Jim Anglin, Chris Jarzynski, Kosuke Shizume, Ben Schumacher, Manny Knill, Jacek Dziarmaga, Diego Dalvit, Zbig Karkuszewski, Harold Ollivier, Roberto Onofrio, Robin Blume-Kohut, David Poulin, Lorenza Viola, and David Wallace.

Finally, I have some advice to the reader. I believe this paper should be read twice: first, just the old text alone; then—and only then—on the second reading, the whole thing. I would also recommend to the curious reader two other overviews: the draft of my *Reviews of Modern Physics* paper (Zurek 2001a) and Les Houches Lectures coauthored with Juan Pablo Paz (Paz and Zurek 2001).

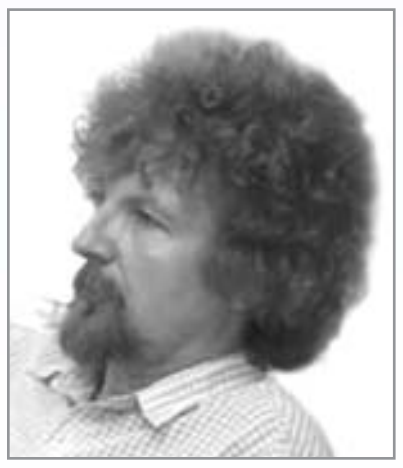

## Introduction

Quantum mechanics works exceedingly well in all practical applications. No example of conflict between its predictions and experiment is known. Without quantum physics, we could not explain the behavior of the solids, the structure and function of DNA, the color of the stars, the action of lasers, or the properties of superfluids. Yet nearly a century after its inception, the debate about the relation of quantum physics to the familiar physical world continues. Why is a theory that seems to account with precision for everything we can measure still deemed lacking?

The only "failure" of quantum theory is its inability to provide a natural framework for our prejudices about the workings of the Universe. States of quantum systems evolve according to the deterministic, linear Schrödinger equation

$$i\hbar \frac{d}{dt}|\psi\rangle = H|\psi\rangle \quad . \tag{1}$$

That is, just as in classical mechanics, given the initial state of the system and its Hamiltonian $H$, one can, at least in principle, compute the state at an arbitrary time. This deterministic evolution of $|\psi\rangle$ has been verified in carefully controlled experiments. Moreover, there is no indication of a border between quantum and classical at which Equation (1) would fail (see cartoon on the opener to this article).

There is, however, a very poorly controlled experiment with results so tangible and immediate that it has enormous power to convince: Our perceptions are often difficult to reconcile with the predictions of Equation (1). Why? Given almost any initial condition, the Universe described by $|\psi\rangle$ evolves into a state containing many alternatives that are never seen to coexist in our world. Moreover, while the ultimate evidence for the choice of one alternative resides in our elusive "consciousness," there is every indication that the choice occurs much before consciousness ever gets involved and that, once made, the choice is irrevocable. Thus, at the root of our unease with quantum theory is the clash between the principle of superposition—the basic tenet of the theory reflected in the linearity of Equation (1)—and everyday classical reality in which this principle appears to be violated.

The problem of measurement has a long and fascinating history. The first widely accepted explanation of how a single outcome emerges from the multitude of potentialities was the Copenhagen Interpretation proposed by Niels Bohr (1928), who insisted that a classical apparatus is necessary to carry out measurements. Thus, quantum theory was not to be universal. The key feature of the Copenhagen Interpretation is the dividing line between quantum and classical. Bohr emphasized that the border must be mobile so that even the "ultimate apparatus"—the human nervous system—could in principle be measured and analyzed as a quantum object, provided that a suitable classical device could be found to carry out the task.

In the absence of a crisp criterion to distinguish between quantum and classical, an identification of the classical with the macroscopic has often been tentatively accepted. The inadequacy of this approach has become apparent as a result of relatively recent developments: A cryogenic version of the Weber bar—a gravity-wave detector— must be treated as a quantum harmonic oscillator even though it may weigh a ton (Braginsky et al. 1980, Caves et al. 1980). Nonclassical squeezed states can describe oscillations of suitably prepared electromagnetic fields with macroscopic numbers of photons (Teich and Saleh 1990). Finally, quantum states associated with the currents of superconducting Josephson junctions involve macroscopic numbers of electrons, but still they can tunnel between the minima of the effective potential corresponding to the opposite sense of rotation (Leggett et al. 1987, Caldeira and Leggett 1983a, Tesche 1986).

If macroscopic systems cannot be always safely placed on the classical side of the boundary, then might there be no boundary at all? The Many Worlds Interpretation (or more accurately, the Many Universes Interpretation), developed by Hugh Everett III with encouragement from John Archibald Wheeler in the 1950s, claims to do away with the boundary (Everett 1957, Wheeler 1957). In this interpretation, the entire universe is described by quantum theory. Superpositions evolve forever according to the Schrödinger equation. Each time a suitable interaction takes place between any two quantum systems, the wave function of the universe splits, developing ever more “branches.”

Initially, Everett’s work went almost unnoticed. It was taken out of mothballs over a decade later by Bryce DeWitt (1970) and DeWitt and Neill Graham (1973), who managed to upgrade its status from “virtually unknown” to “very controversial.” The Many Worlds Interpretation is a natural choice for quantum cosmology, which describes the whole Universe by means of a state vector. There is nothing more macroscopic than the Universe. It can have no a priori classical subsystems. There can be no observer “on the outside.” In this universal setting, classicality must be an emergent property of the selected observables or systems.

At first glance, the Many Worlds and Copenhagen Interpretations have little in common. The Copenhagen Interpretation demands an a priori “classical domain” with a border that enforces a classical “embargo” by letting through just one potential outcome. The Many Worlds Interpretation aims to abolish the need for the border altogether. Every potential outcome is accommodated by the ever-proliferating branches of the wave function of the Universe. The similarity between the difficulties faced by these two viewpoints becomes apparent, nevertheless, when we ask the obvious question, “Why do I, the observer, perceive only one of the outcomes?” Quantum theory, with its freedom to rotate bases in Hilbert space, does not even clearly define which states of the Universe correspond to the “branches.” Yet, our perception of a reality with alternatives—not a coherent superposition of alternatives—demands an explanation of when, where, and how it is decided what the observer actually records. Considered in this context, the Many Worlds Interpretation in its original version does not really abolish the border but pushes it all the way to the boundary between the physical Universe and consciousness. Needless to say, this is a very uncomfortable place to do physics.

In spite of the profound nature of the difficulties, recent years have seen a growing consensus that progress is being made in dealing with the measurement problem, which is the usual euphemism for the collection of interpretational conundrums described above. The key (and uncontroversial) fact has been known almost since the inception of quantum theory, but its significance for the transition from quantum to classical is being recognized only now: Macroscopic systems are never isolated from their environments. Therefore—as H. Dieter Zeh emphasized (1970)—they should not be expected to follow Schrödinger’s equation, which is applicable only to a closed system. As a result, systems usually regarded as classical suffer (or benefit) from the natural loss of quantum coherence, which “leaks out” into the environment (Zurek 1981, 1982). The resulting “decoherence” cannot be ignored when one addresses the problem of the reduction of the quantum mechanical wave packet: Decoherence imposes, in effect, the required “embargo” on the potential outcomes by allowing the observer to maintain records of alternatives but to be aware of only one of the branches—one of the “decoherent histories” in the nomenclature of Murray Gell-Mann and James Hartle (1990) and Hartle (1991).

The aim of this paper is to explain the physics and thinking behind this approach. The reader should be warned that this writer is not a disinterested witness to this development (Wigner 1983, Joos and Zeh 1985, Haake and Walls 1986, Milburn and Holmes 1986, Albrecht 1991, Hu et al. 1992), but rather, one of the proponents. I shall, nevertheless, attempt to paint a fairly honest picture and point out the difficulties as well as the accomplishments.

### Decoherence in Quantum Information Processing

Much of what was written in the introduction remains valid today. One important development is the increase in experimental evidence for the validity of the quantum principle of superposition in various contexts including spectacular double-slit experiments that demonstrate interference of fullerenes (Arndt et al. 1999), the study of superpositions in Josephson junctions (Mooij et al.1999, Friedman et al. 2000), and the implementation of Schrödinger's "kittens" in atom interferometry (Chapman et al. 1995, Pfau et al. 1994), ion traps (Monroe et al. 1996) and microwave cavities (Brune et al. 1996). In addition to confirming the superposition principle and other exotic aspects of quantum theory (such as entanglement) in novel settings, these experiments allow—as we shall see later—for a controlled investigation of decoherence.

The other important change that influenced the perception of the quantum-to-classical "border territory" is the explosion of interest in quantum information and computation. Although quantum computers were already being discussed in the 1980s, the nature of the interest has changed since Peter Shor invented his factoring algorithm. Impressive theoretical advances, including the discovery of quantum error correction and resilient quantum computation, quickly followed, accompanied by increasingly bold experimental forays. The superposition principle, once the cause of trouble for the interpretation of quantum theory, has become the central article of faith in the emerging science of quantum information processing. This last development is discussed elsewhere in this issue, so I shall not dwell on it here.

The application of quantum physics to information processing has also transformed the nature of interest in the process of decoherence: At the time of my original review (1991), decoherence was a solution to the interpretation problem—a mechanism to impose an effective classicality on de facto quantum systems. In quantum information processing, decoherence plays two roles. Above all, it is a threat to the quantumness of quantum information. It invalidates the quantum superposition principle and thus turns quantum computers into (at best) classical computers, negating the potential power offered by the quantumness of the algorithms. But decoherence is also a necessary (although often taken for granted) ingredient in quantum information processing, which must, after all, end in a "measurement."

The role of a measurement is to convert quantum states and quantum correlations (with their characteristic indefiniteness and malleability) into classical, definite outcomes. Decoherence leads to the environment-induced superselection (einselection) that justifies the existence of the preferred pointer states. It enables one to draw an effective border between the quantum and the classical in straightforward terms, which do not appeal to the "collapse of the wave packet" or any other such deus ex machina.

## Correlations and Measurements

A convenient starting point for the discussion of the measurement problem and, more generally, of the emergence of classical behavior from quantum dynamics is the analysis of quantum measurements due to John von Neumann (1932). In contrast to Bohr, who assumed at the outset that the apparatus must be classical (thereby forfeiting claims that quantum theory is universal), von Neumann analyzed the case of a quantum apparatus. I shall reproduce his analysis for the simplest case: a measurement on a two-state system $\mathcal{S}$ (which can be thought of as an atom with spin 1/2) in which a quantum two-state (one bit) detector records the result.

The Hilbert space $\mathcal{H}_\mathcal{S}$ of the system is spanned by the orthonormal states $|\uparrow\rangle$ and $|\downarrow\rangle$, while the states $|d_\uparrow\rangle$ and $|d_\downarrow\rangle$ span the $\mathcal{H}_\mathcal{D}$ of the detector. A two-dimensional $\mathcal{H}_\mathcal{D}$ is the absolute minimum needed to record the possible outcomes. One can devise a quantum

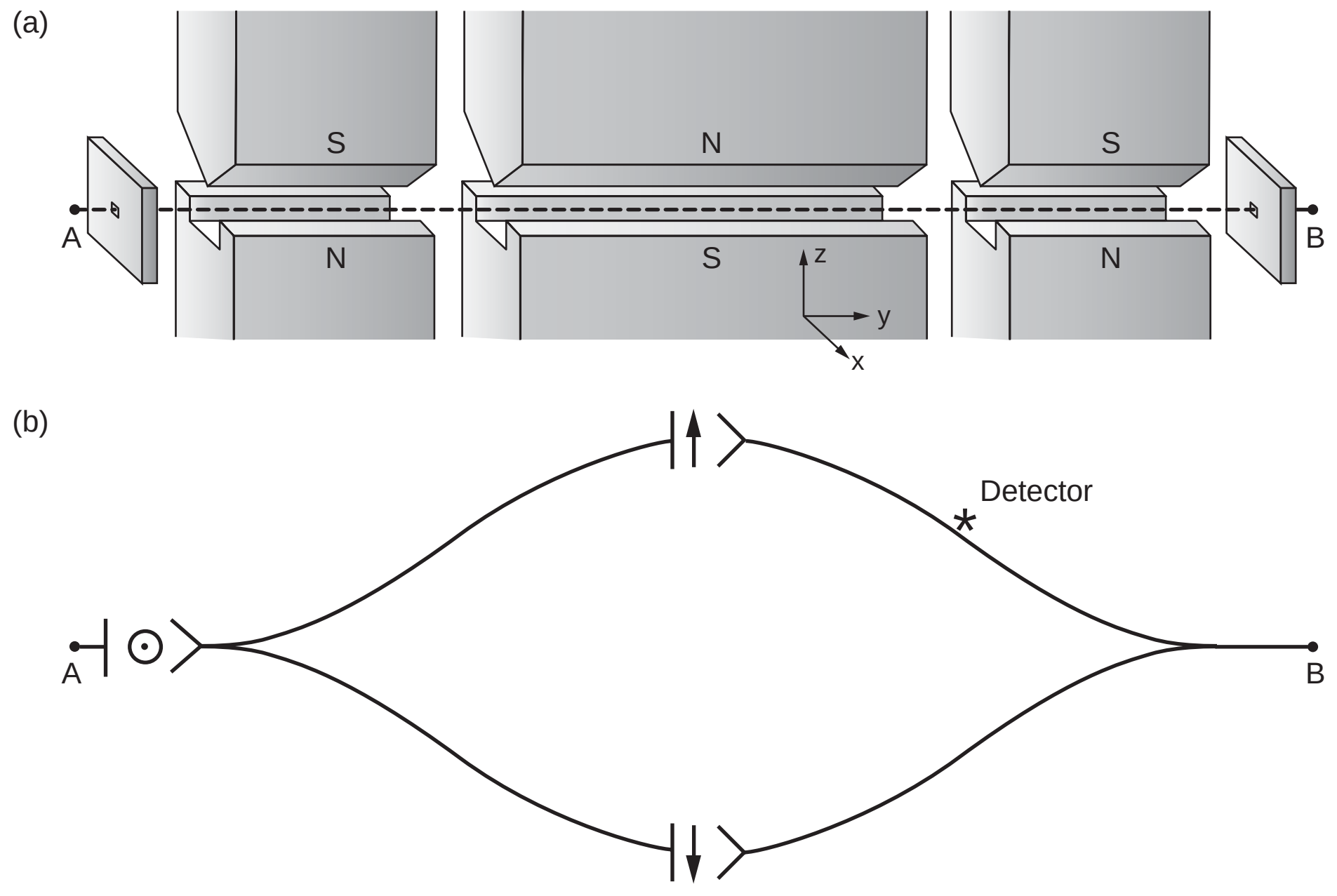

**Figure 1. A Reversible Stern-Gerlach Apparatus**
**The "gedanken" reversible Stern-Gerlach apparatus in (a) splits a beam of atoms into two branches that are correlated with the component of the spin of the atoms (b) and then recombines the branches before the atoms leave the device. Eugene Wigner (1963) used this gedanken experiment to show that a correlation between the spin and the location of an atom can be reversibly undone. The introduction of a one-bit (two-state) quantum detector that changes its state when the atom passes nearby prevents the reversal: The detector inherits the correlation between the spin and the trajectory, so the Stern-Gerlach apparatus can no longer undo the correlation.** (This illustration was adapted with permission from Zurek 1981.)

detector (see Figure 1) that "clicks" only when the spin is in the state $|\uparrow\rangle$, that is,

$$|\uparrow\rangle\,|d_\downarrow\rangle \rightarrow |\uparrow\rangle\,|d_\uparrow\rangle \quad , \tag{2}$$

and remains unperturbed otherwise (Zeh 1970, Wigner 1963, Scully et al. 1989).

I shall assume that, before the interaction, the system was in a pure state $|\psi_{\mathcal{S}}\rangle$ given by

$$|\psi_{\mathcal{S}}\rangle = \alpha|\uparrow\rangle + \beta|\downarrow\rangle \quad , \tag{3}$$

with the complex coefficients satisfying $|\alpha|^2 + |\beta|^2 = 1$. The composite system starts as

$$|\Phi^i\rangle = |\psi_{\mathcal{S}}\rangle|d_\downarrow\rangle \quad . \tag{4}$$

Interaction results in the evolution of $|\Phi^i\rangle$ into a correlated state $|\Phi^c\rangle$:

$$|\Phi^i\rangle = (\alpha|\uparrow\rangle + \beta|\downarrow\rangle)|d_\downarrow\rangle \Rightarrow \alpha|\uparrow\rangle|d_\uparrow\rangle + \beta|\downarrow\rangle|d_\downarrow\rangle = |\Phi^c\rangle \quad . \tag{5}$$

This essential and uncontroversial first stage of the measurement process can be accomplished by means of a Schrödinger equation with an appropriate interaction. It might be tempting to halt the discussion of measurements with Equation (5). After all, the correlated state vector $|\Phi^c\rangle$ implies that, if the detector is seen in the state $|d_\uparrow\rangle$, the system is guaranteed to be found in the state $|\uparrow\rangle$. Why ask for anything more?

The reason for dissatisfaction with $|\Phi^c\rangle$ as a description of a completed measurement is simple and fundamental: In the real world, even when we do not know the outcome of a measurement, we do know the possible alternatives, and we can safely act as if only one of those alternatives has occurred. As we shall see in the next section, such an assumption is not only unsafe but also simply wrong for a system described by $|\Phi^c\rangle$.

How then can an observer (who has not yet consulted the detector) express his ignorance about the outcome without giving up his certainty about the "menu" of the possibilities? Quantum theory provides the right formal tool for the occasion: A density matrix can be used to describe the probability distribution over the alternative outcomes.

Von Neumann was well aware of these difficulties. Indeed, he postulated (1932) that, in addition to the unitary evolution given by Equation (1), there should be an ad hoc "process 1"—a nonunitary reduction of the state vector—that would take the pure, correlated state $|\Phi^C\rangle$ into an appropriate mixture: This process makes the outcomes independent of one another by taking the pure-state density matrix:

$$\rho^c = |\Phi^c\rangle\langle\Phi^c| = |\alpha|^2|\uparrow\rangle\langle\uparrow||d_\uparrow\rangle\langle d_\uparrow| + \alpha\beta^*|\uparrow\rangle\langle\downarrow||d_\uparrow\rangle\langle d_\downarrow|$$

$$+ \alpha^*\beta|\downarrow\rangle\langle\uparrow|d_\downarrow\rangle\langle d_\uparrow| + |\beta|^2|\downarrow\rangle\langle\downarrow||d_\downarrow\rangle\langle d_\downarrow| \quad , \tag{6}$$

and canceling the off-diagonal terms that express purely quantum correlations (entanglement) so that the reduced density matrix with only classical correlations emerges:

$$\rho^r = |\alpha|^2|\uparrow\rangle\langle\uparrow||d_\uparrow\rangle\langle d_\uparrow| + |\beta|^2|\downarrow\rangle\langle\downarrow||d_\downarrow\rangle\langle d_\downarrow| \quad . \tag{7}$$

Why is the reduced $\rho^r$ easier to interpret as a description of a completed measurement than $\rho^c$? After all, both $\rho^r$ and $\rho^c$ contain identical diagonal elements. Therefore, both outcomes are still potentially present. So what—if anything—was gained at the substantial price of introducing a nonunitary process 1?

## The Question of Preferred Basis: What Was Measured?

The key advantage of $\rho^r$ over $\rho^c$ is that its coefficients may be interpreted as classical probabilities. The density matrix $\rho^r$ can be used to describe the alternative states of a composite spin-detector system that has classical correlations. Von Neumann's process 1 serves a similar purpose to Bohr's "border" even though process 1 leaves all the alternatives in place. When the off-diagonal terms are absent, one can nevertheless safely maintain that the apparatus, as well as the system, is each separately in a definite but unknown state, and that the correlation between them still exists in the preferred basis defined by the states appearing on the diagonal. By the same token, the identities of two halves of a split coin placed in two sealed envelopes may be unknown but are classically correlated. Holding one unopened envelope, we can be sure that the half it contains is either "heads" or "tails" (and not some superposition of the two) and that the second envelope contains the matching alternative.

By contrast, it is impossible to interpret $\rho^c$ as representing such "classical ignorance." In particular, even the set of the alternative outcomes is not decided by $\rho^c$! This circumstance can be illustrated in a dramatic fashion by choosing $\alpha = -\beta = 1/\sqrt{2}$ so that the density matrix $\rho^c$ is a projection operator constructed from the correlated state

$$|\Phi^C\rangle = (|\uparrow\rangle|d_\uparrow\rangle - |\downarrow\rangle|d_\downarrow\rangle)/\sqrt{2} \quad . \tag{8}$$

This state is invariant under the rotations of the basis. For instance, instead of the eigenstates of $|\uparrow\rangle$ and $|\downarrow\rangle$ of $\hat{\sigma}_z$ one can rewrite $|\Phi^C\rangle$ in terms of the eigenstates of $\hat{\sigma}_x$:

$$|\odot\rangle = (|\uparrow\rangle + |\downarrow\rangle)/\sqrt{2} \quad , \tag{9a}$$

$$|\otimes\rangle = (|\uparrow\rangle - |\downarrow\rangle)/\sqrt{2} \quad . \tag{9b}$$

This representation immediately yields

$$|\Phi^C\rangle = -(|\odot\rangle|d_\odot\rangle - |\otimes\rangle|d_\otimes\rangle)/\sqrt{2} \quad , \tag{10}$$

where

$$|d_\odot\rangle = (|d_\downarrow\rangle - |d_\uparrow\rangle)/\sqrt{2} \text{ and } |d_\otimes\rangle = (|d_\uparrow\rangle + |d_\downarrow\rangle)/\sqrt{2} \tag{11}$$

are, as a consequence of the superposition principle, perfectly "legal" states in the Hilbert space of the quantum detector. Therefore, the density matrix

$$\rho^c = |\Phi^C\rangle\langle\Phi^C|$$

could have many (in fact, infinitely many) different states of the subsystems on the diagonal.

This freedom to choose a basis should not come as a surprise. Except for the notation, the state vector $|\Phi^c\rangle$ is the same as the wave function of a pair of maximally correlated (or entangled) spin-1/2 systems in David Bohm's version (1951) of the Einstein-Podolsky-Rosen (EPR) paradox (Einstein et al. 1935). And the experiments that show that such nonseparable quantum correlations violate Bell's inequalities (Bell 1964) are demonstrating the following key point: The states of the two spins in a system described by $|\Phi^c\rangle$ are not just unknown, but rather they cannot exist before the "real" measurement (Aspect et al. 1981, 1982). We conclude that when a detector is quantum, a superposition of records exists and is a record of a superposition of outcomes—a very nonclassical state of affairs.

## Missing Information and Decoherence

Unitary evolution condemns every closed quantum system to "purity." Yet, if the outcomes of a measurement are to become independent events, with consequences that can be explored separately, a way must be found to dispose of the excess information and thereby allow any orthogonal basis—any potential events and their superpositions—to be equally correlated. In the previous sections, quantum correlation was analyzed from the point of view of its role in acquiring information. Here, I shall discuss the flip side of the story: Quantum correlations can also disperse information throughout the degrees of freedom that are, in effect, inaccessible to the observer. Interaction with the degrees of freedom external to the system—which we shall summarily refer to as the environment—offers such a possibility.

Reduction of the state vector, $\rho^c \Rightarrow \rho^r$, decreases the information available to the observer about the composite system $\mathcal{SD}$. The information loss is needed if the outcomes are to become classical and thereby available as initial conditions to predict the future. The effect of this loss is to increase the entropy $\mathcal{H} = -Tr\rho \lg\rho$ by an amount

$$\Delta\mathcal{H} = \mathcal{H}(\rho^r) - \mathcal{H}(\rho^c) = -(|\alpha|^2 \lg|\alpha|^2 + |\beta|^2 \lg|\beta|^2) \quad . \tag{12}$$

Entropy must increase because the initial state described by $\rho^c$ was pure, $\mathcal{H}(\rho^c) = 0$, and the reduced state is mixed. Information gain—the objective of the measurement—is accomplished only when the observer interacts and becomes correlated with the detector in the already precollapsed state $\rho^r$.

To illustrate the process of the environment-induced decoherence, consider a system $\mathcal{S}$, a detector $\mathcal{D}$, and an environment $\mathcal{E}$. The environment is also a quantum system. Following the first step of the measurement process—establishment of a correlation as shown in Equation (5)—the environment similarly interacts and becomes correlated with the apparatus:

$$|\Phi^c\rangle|\mathcal{E}_0\rangle = (\alpha|\uparrow\rangle|d_\uparrow\rangle + \beta|\downarrow\rangle|d_\downarrow\rangle)|\mathcal{E}_0\rangle \Rightarrow \alpha|\uparrow\rangle|d_\uparrow\rangle|\mathcal{E}_\uparrow\rangle + \beta|\downarrow\rangle|d_\downarrow\rangle|\mathcal{E}_\downarrow\rangle = |\Psi\rangle \ . \qquad (13)$$

The final state of the combined $\mathcal{SDE}$ "von Neumann chain" of correlated systems extends the correlation beyond the $\mathcal{SD}$ pair. When the states of the environment $|\mathcal{E}_i\rangle$ corresponding to the states $|d_\uparrow\rangle$ and $|d_\downarrow\rangle$ of the detector are orthogonal, $\langle\mathcal{E}_i|\mathcal{E}_{i'}\rangle = \delta_{ii'}$, the density matrix for the detector-system combination is obtained by ignoring (tracing over) the information in the uncontrolled (and unknown) degrees of freedom

$$\rho_{\mathcal{DS}} = Tr_{\mathcal{E}}\,|\Psi\rangle\langle\Psi| = \sum_i \langle\mathcal{E}_i|\Psi\rangle\langle\Psi|\mathcal{E}_{i'}\rangle = |\alpha|^2|\uparrow\rangle\langle\uparrow||d_\uparrow\rangle\langle d_\uparrow| + |\beta|^2|\downarrow\rangle\langle\downarrow||d_\downarrow\rangle\langle d_\downarrow| = \rho^r \ . \qquad (14)$$

The resulting $\rho^r$ is precisely the reduced density matrix that von Neumann called for. Now, in contrast to the situation described by Equations (9)–(11), a superposition of the records of the detector states is no longer a record of a superposition of the state of the system. A preferred basis of the detector, sometimes called the "pointer basis" for obvious reasons, has emerged. Moreover, we have obtained it—or so it appears—without having to appeal to von Neumann's nonunitary process 1 or anything else beyond the ordinary, unitary Schrödinger evolution. The preferred basis of the detector—or for that matter, of any open quantum system—is selected by the dynamics.

Not all aspects of this process are completely clear. It is, however, certain that the detector–environment interaction Hamiltonian plays a decisive role. In particular, when the interaction with the environment dominates, eigenspaces of any observable $\Lambda$ that commutes with the interaction Hamiltonian,

$$[\Lambda, H_{int}] = 0 \ , \qquad (15)$$

invariably end up on the diagonal of the reduced density matrix (Zurek 1981, 1982). This commutation relation has a simple physical implication: It guarantees that the pointer observable $\Lambda$ will be a constant of motion, a conserved quantity under the evolution generated by the interaction Hamiltonian. Thus, when a system is in an eigenstate of $\Lambda$, interaction with the environment will leave it unperturbed.

In the real world, the spreading of quantum correlations is practically inevitable. For example, when in the course of measuring the state of a spin-1/2 atom (see Figure 1b), a photon had scattered from the atom while it was traveling along one of its two alternative routes, this interaction would have resulted in a correlation with the environment and would have necessarily led to a loss of quantum coherence. The density matrix of the $\mathcal{SD}$ pair would have lost its off-diagonal terms. Moreover, given that it is impossible to catch up with the photon, such loss of coherence would have been irreversible. As we shall see later, irreversibility could also arise from more familiar, statistical causes: Environments are notorious for having large numbers of interacting degrees of freedom, making extraction of lost information as difficult as reversing trajectories in the Boltzmann gas.

## Quantum Discord—A Measure of Quantumness

The contrast between the density matrices in Equations (6) and (7) is stark and obvious. In particular, the entanglement between the system and the detector in $\rho^c$ is obviously quantum—classical systems cannot be entangled. The argument against the "ignorance" interpretation of $\rho^c$ still stands. Yet we would like to have a quantitative measure of how much is classical (or how much is quantum) about the correlations of a state represented by a general density matrix. Such a measure of the quantumness of correlation was devised recently (Ollivier and Zurek 2002). It is known as quantum discord. Of the several closely related definitions of discord, we shall select one that is easiest to explain. It is based on mutual information—an information-theoretic measure of how much easier it is to describe the state of a pair of objects $(\mathcal{S}, \mathcal{D})$ jointly rather than separately. One formula for mutual information $\mathcal{I}(\mathcal{S}:\mathcal{D})$ is simply

$$\mathcal{I}(\mathcal{S}:\mathcal{D}) = \mathcal{H}(\mathcal{S}) + \mathcal{H}(\mathcal{D}) - \mathcal{H}(\mathcal{S}, \mathcal{D}),$$

where $\mathcal{H}(\mathcal{S})$ and $\mathcal{H}(\mathcal{D})$ are the entropies of $\mathcal{S}$ and $\mathcal{D}$, respectively, and $\mathcal{H}(\mathcal{S}, \mathcal{D})$ is the joint entropy of the two. When $\mathcal{S}$ and $\mathcal{D}$ are not correlated (statistically independent),

$$\mathcal{H}(\mathcal{S}, \mathcal{D}) = \mathcal{H}(\mathcal{S}) + \mathcal{H}(\mathcal{D}),$$

and $\mathcal{I}(\mathcal{S}:\mathcal{D}) = 0$. By contrast, when there is a perfect classical correlation between them (for example, two copies of the same book), $\mathcal{H}(\mathcal{S}, \mathcal{D}) = \mathcal{H}(\mathcal{S}) = \mathcal{H}(\mathcal{D}) = \mathcal{I}(\mathcal{S}:\mathcal{D})$. Perfect classical correlation implies that, when we find out all about one of them, we also know everything about the other, and the conditional entropy $\mathcal{H}(\mathcal{S}|\mathcal{D})$ (a measure of the uncertainty about $\mathcal{S}$ after the state of $\mathcal{D}$ is found out) disappears. Indeed, classically, the joint entropy $\mathcal{H}(\mathcal{S}, \mathcal{D})$ can always be decomposed into, say $\mathcal{H}(\mathcal{D})$, which measures the information missing about $\mathcal{D}$, and the conditional entropy $\mathcal{H}(\mathcal{S}|\mathcal{D})$. Information is still missing about $\mathcal{S}$ even after the state of $\mathcal{D}$ has been determined: $\mathcal{H}(\mathcal{S}, \mathcal{D}) = \mathcal{H}(\mathcal{D}) + \mathcal{H}(\mathcal{S}|\mathcal{D})$. This expression for the joint entropy suggests an obvious rewrite of the preceding definition of mutual information into a classically identical form, namely,

$$\mathcal{J}(\mathcal{S}:\mathcal{D}) = \mathcal{H}(\mathcal{S}) + \mathcal{H}(\mathcal{D}) - (\mathcal{H}(\mathcal{D}) + \mathcal{H}(\mathcal{S}|\mathcal{D})).$$

Here, we have abstained from the obvious (and perfectly justified from a classical viewpoint) cancellation in order to emphasize the central feature of quantumness: In quantum physics, the state collapses into one of the eigenstates of the measured observable. Hence, a state of the object is redefined by a measurement. Thus, the joint entropy can be defined in terms of the conditional entropy only after the measurement used to access, say, $\mathcal{D}$, has been specified. In that case,

$$\mathcal{H}_{|d_k\rangle}(\mathcal{S}, \mathcal{D}) = (\mathcal{H}(\mathcal{D}) + \mathcal{H}(\mathcal{S}|\mathcal{D}))_{|d_k\rangle}\ .$$

This type of joint entropy expresses the ignorance about the pair $(\mathcal{S}, \mathcal{D})$ after the observable with the eigenstates $\{|d_k\rangle\}$ has been measured on $\mathcal{D}$. Of course, $\mathcal{H}_{|d_k\rangle}(\mathcal{S}, \mathcal{D})$ is not the only way to define the entropy of the pair. One can also compute a basis-independent joint entropy $\mathcal{H}(\mathcal{S}, \mathcal{D})$, the von Neumann entropy of the pair. Since these two definitions of joint entropy do not coincide in the quantum case, we can define a basis-dependent quantum discord

$$\delta_{|d_k\rangle}(\mathcal{S}|\mathcal{D}) = \mathcal{I} - \mathcal{J} = (\mathcal{H}(\mathcal{D}) + \mathcal{H}(\mathcal{S}|\mathcal{D}))_{|d_k\rangle} - \mathcal{H}(\mathcal{S},\mathcal{D})$$

as the measure of the extent by which the underlying density matrix describing $\mathcal{S}$ and $\mathcal{D}$ is perturbed by a measurement of the observable with the eigenstates $\{|d_k\rangle\}$. States of classical objects—or classical correlations—are "objective:" They exist independent of measurements. Hence, when there is a basis $\{|\hat{d}_k\rangle\}$ such that the minimum discord evaluated for this basis disappears,

$$\hat{\delta}(\mathcal{S}|\mathcal{D}) = \min_{\{|d_k\rangle\}}\big(\mathcal{H}(\mathcal{S},\mathcal{D}) - (\mathcal{H}(\mathcal{D}) + \mathcal{H}(\mathcal{S}|\mathcal{D})\big)_{|d_k\rangle} = 0,$$

the correlation can be regarded as effectively classical (or more precisely, as "classically accessible through $\mathcal{D}$"). One can then show that there is a set of probabilities associated with the basis $\{|d_k\rangle\}$ that can be treated as classical. It is straightforward to see that, when $\mathcal{S}$ and $\mathcal{D}$ are entangled (for example, $\rho^c = |\phi^c\rangle\langle\phi^c|$), then $\hat{\delta} > 0$ in all bases. By contrast, if we consider $\rho^r$, discord disappears in the basis $\{|d_\uparrow\rangle, |d_\downarrow\rangle\}$ so that the underlying correlation is effectively classical.

It is important to emphasize that quantum discord is not just another measure of entanglement but a genuine measure of the quantumness of correlations. In situations involving measurements and decoherence, quantumness disappears for the preferred set of states that are effectively classical and thus serves as an indicator of the pointer basis, which as we shall see, emerges as a result of decoherence and einselection.

## Decoherence: How Long Does It Take?

A tractable model of the environment is afforded by a collection of harmonic oscillators (Feynman and Vernon 1963, Dekker 1981, Caldeira and Leggett 1983a, 1983b, 1985, Joos and Zeh 1985, Paz et al. 1993) or, equivalently, by a quantum field (Unruh and Zurek 1989). If a particle is present, excitations of the field will scatter off the particle. The resulting "ripples" will constitute a record of its position, shape, orientation, and so on, and most important, its instantaneous location and hence its trajectory.

A boat traveling on a quiet lake or a stone that fell into water will leave such an imprint on the water surface. Our eyesight relies on the perturbation left by the objects on the preexisting state of the electromagnetic field. Hence, it is hardly surprising that an imprint is left whenever two quantum systems interact, even when "nobody is looking," and even when the lake is stormy and full of preexisting waves, and the field is full of excitations—that is, when the environment starts in equilibrium at some finite temperature. "Messy" initial states of the environment make it difficult to decipher the record, but do not preclude its existence.

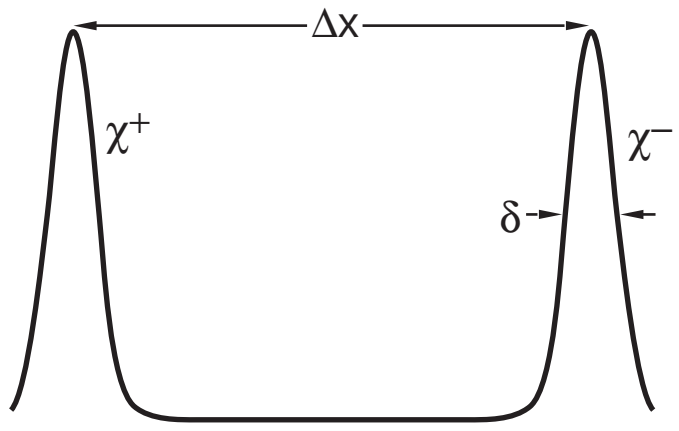

**Figure 2. A "Schrödinger Cat" State or a Coherent Superposition**
**This cat state $\varphi(x)$, the coherent superposition of two Gaussian wave packets of Equation (18), could describe a particle in a superposition of locations inside a Stern-Gerlach apparatus (see Figure 1) or the state that develops in the course of a double-slit experiment. The phase between the two components has been chosen to be zero.**

A specific example of decoherence—a particle at position $x$ interacting with a scalar field $\varphi$ (which can be regarded as a collection of harmonic oscillators) through the Hamiltonian

$$H_{int} = \epsilon x \, d\phi/dt \tag{16}$$

has been extensively studied by many, including the investigators just referenced. The conclusion is easily formulated in the so-called "high-temperature limit," in which only thermal-excitation effects of the field $\phi$ are taken into account and the effect of zero-point vacuum fluctuations is neglected.

In this case, the density matrix $\rho(x, x')$ of the particle in the position representation evolves according to the master equation

$$\dot{\rho} = \overbrace{\underbrace{-\frac{i}{\hbar}[H,\rho]}_{\dot{p}=-FORCE=\nabla V}}^{Von\ Neumann\ Equation} - \overbrace{\underbrace{\gamma(x-x')\left(\frac{\partial}{\partial x}-\frac{\partial}{\partial x'}\right)\rho}_{\dot{p}=-\gamma p}}^{Relaxation} - \overbrace{\underbrace{\frac{2m\gamma k_B T}{\hbar^2}(x-x')^2\rho}_{Classical\ Phase\ Space}}^{Decoherence}, \tag{17}$$

where $H$ is the particle's Hamiltonian (although with the potential $V(x)$ adjusted because of $H_{int}$), $\gamma$ is the relaxation rate, $k_B$ is the Boltzmann constant, and $T$ is the temperature of the field. Equation (17) is obtained by first solving exactly the Schrödinger equation for a particle plus the field and then tracing over the degrees of freedom of the field.

I will not analyze Equation (17) in detail but just point out that it naturally separates into three distinct terms, each of them responsible for a different aspect of the effectively classical behavior. The first term—the von Neumann equation (which can be derived from the Schrödinger equation)—generates reversible classical evolution of the expectation value of any observable that has a classical counterpart regardless of the form of $\rho$ (Ehrenfest's theorem). The second term causes dissipation. The relaxation rate $\gamma = \eta/2m$ is proportional to the viscosity $\eta = \epsilon^2/2$ due to the interaction with the scalar field. That interaction causes a decrease in the average momentum and loss of energy. The last term also has a classical counterpart: It is responsible for fluctuations or random "kicks" that lead to Brownian motion. We shall see this in more detail in the next section.

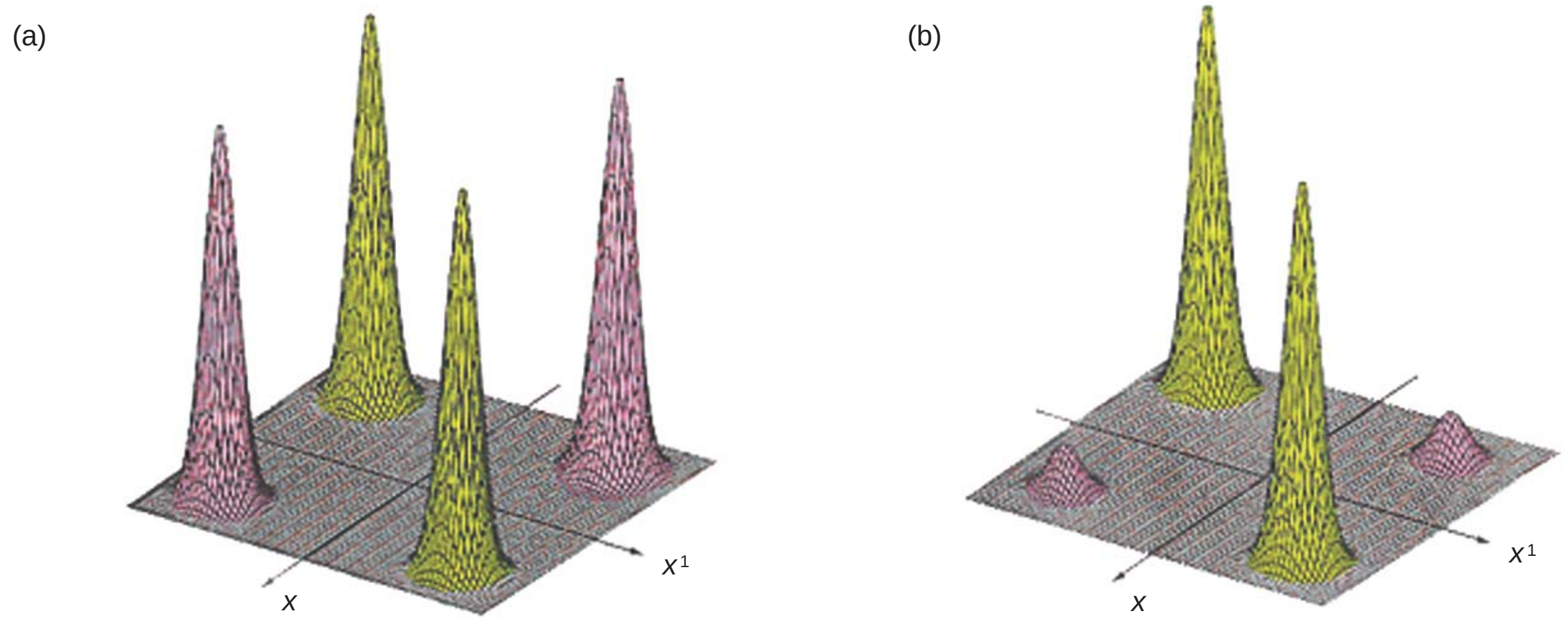

**Figure 3. Evolution of the Density Matrix for the Schrödinger Cat State in Figure 2**

**(a)This plot shows the density matrix for the cat state in Figure 2 in the position representation $\rho(x, x') = \varphi(x)\varphi^*(x)$. The peaks near the diagonal (green) correspond to the two possible locations of the particle. The peaks away from the diagonal (red) are due to quantum coherence. Their existence and size demonstrate that the particle is not in either of the two approximate locations but in a coherent superposition of them. (b) Environment-induced decoherence causes decay of the off-diagonal terms of $\rho(x, x')$. Here, the density matrix in (a) has partially decohered. Further decoherence would result in a density matrix with diagonal peaks only. It can then be regarded as a classical probability distribution with an equal probability of finding the particle in either of the locations corresponding to the Gaussian wave packets.**

For our purposes, the effect of the last term on quantum superpositions is of greatest interest. I shall show that it destroys quantum coherence, eliminating off-diagonal terms responsible for quantum correlations between spatially separated pieces of the wave packet. It is therefore responsible for the classical structure of the phase space, as it converts superpositions into mixtures of localized wave packets which, in the classical limit, turn into the familiar points in phase space. This effect is best illustrated by an example. Consider the "cat" state shown in Figure 2, where the wave function of a particle is given by a coherent superposition of two Gaussians: $\varphi(x) = (\chi^+(x) + \chi^-(x))/2^{1/2}$ and the Gaussians are

$$\chi^{\pm}(x) = \langle x|\pm\rangle \sim \exp\left[-\frac{\left(x \pm \frac{\Delta x}{2}\right)^2}{4\delta^2}\right]. \tag{18}$$

For the case of wide separation ($\Delta x >> \delta$), the corresponding density matrix $\rho(x, x') = \varphi(x)\,\varphi^*(x')$ has four peaks: Two on the diagonal defined by $x = x'$, and two off the diagonal for which $x$ and $x'$ are very different (see Figure 3). Quantum coherence is due to the off-diagonal peaks. As those peaks disappear, position emerges as an approximate preferred basis.

The last term of Equation (17), which is proportional to $(x - x')^2$, has little effect on the diagonal peaks. By contrast, it has a large effect on the off-diagonal peaks for which $(x - x')^2$ is approximately the square of the separation $(\Delta x)^2$. In particular, it causes the off-diagonal peaks to decay at the rate $\frac{d}{dt}\left(\rho^{+-}\right) \sim 2\gamma m k_B T \big/ \hbar^2 \left(\Delta x\right)^2 \rho^{+-} = \tau_D^{-1}\rho^{+}$ .

It follows that quantum coherence will disappear on a decoherence time scale (Zurek 1984).

$$\tau_D \cong \gamma^{-1}\left(\frac{\lambda_{dB}}{\Delta x}\right)^2 = \tau_R\left(\frac{\hbar}{\Delta x\sqrt{2mk_BT}}\right)^2, \tag{19}$$

where $\lambda_{dB} = \hbar/(2mk_BT)^{-1/2}$ is the thermal de Broglie wavelength. For macroscopic objects, the decoherence time $\tau_D$ is typically much less than the relaxation time $\tau_R = \gamma^{-1}$.

For a system at temperature $T$ = 300 kelvins with mass $m$ = 1 gram and separation $\Delta x$ = 1 centimeter, the ratio of the two time scales is $\tau_D/\tau_R \sim 10^{-40}$! Thus, even if the relaxation rate were of the order of the age of the Universe, ~$10^{17}$ seconds, quantum coherence would be destroyed in $\tau_D \sim 10^{-23}$ second.

For microscopic systems and, occasionally, even for very macroscopic ones, the decoherence times are relatively long. For an electron ($m_e = 10^{-27}$ grams), $\tau_D$ can be much larger than the other relevant time scales on atomic and larger energy and distance scales. For a massive Weber bar, tiny $\Delta x$ (~$10^{-17}$ centimeter) and cryogenic temperatures suppress decoherence. Nevertheless, the macroscopic nature of the object is certainly crucial in facilitating the transition from quantum to classical.

### Experiments on Decoherence

A great deal of work on master equations and their derivations in different situations has been conducted since 1991, but in effect, most of the results described above stand. A summary can be found in Paz and Wojchiech Zurek (2001**)** and a discussion of the caveats to the simple conclusions regarding decoherence rates appears in Anglin et al. (1997).

Perhaps the most important development in the study of decoherence is on the experimental front. In the past decade, several experiments probing decoherence in various systems have been carried out. In particular, Michel Brune, Serge Haroche, Jean-Michel Raimond, and their colleagues at École Normale Supérieure in Paris (Brune et al. 1996, Haroche 1998) have performed a series of microwave cavity experiments in which they manipulate electromagnetic fields into a Schrödinger-cat-like superposition using rubidium atoms. They probe the ensuing loss of quantum coherence. These experiments have confirmed the basic tenets of decoherence theory. Since then, the French scientists have applied the same techniques to implement various quantum information-processing ventures. They are in the process of upgrading their equipment in order to produce “bigger and better” Schrödinger cats and to study their decoherence.

A little later, Wineland, Monroe, and coworkers (Turchette et al. 2000) used ion traps (set up to implement a fragment of one of the quantum computer designs) to study the decoherence of ions due to radiation. Again, theory was confirmed, further advancing the status of decoherence as both a key ingredient of the explanation of the emergent classicality and a threat to quantum computation. In addition to these developments, which test various aspects of decoherence induced by a real or simulated “large environment,” Pritchard and his coworkers at the Massachusetts Institute of Technology have carried out a beautiful sequence of experiments by using atomic interferometry in order to investigate the role of information transfer between atoms and photons (see Kokorowski et al. 2001 and other references therein). Finally, “analogue experiments” simulating the behavior of the Schrödinger equation in optics (Cheng and Raymer 1999) have explored some of the otherwise difficult-to-access corners of the parameter space.

In addition to these essentially mesoscopic Schrödinger-cat decoherence experiments, designs of much more substantial “cats” (for example, mirrors in superpositions of quantum states) are being investigated in several laboratories.

## Classical Limit of Quantum Dynamics

The Schrödinger equation was deduced from classical mechanics in the Hamilton-Jacobi form. Thus, it is no surprise that it yields classical equations of motion when $\hbar$ can be regarded as small. This fact, along with Ehrenfest's theorem, Bohr's correspondence principle, and the kinship of quantum commutators with the classical Poisson brackets, is part of the standard lore found in textbooks. However, establishing the quantum–classical correspondence involves the states as well as the equations of motion. Quantum mechanics is formulated in Hilbert space, which can accommodate localized wave packets with sensible classical limits as well as the most bizarre superpositions. By contrast, classical dynamics happens in phase space.

To facilitate the study of the transition from quantum to classical behavior, it is convenient to employ the Wigner transform of a wave function $\psi(x)$:

$$W(x,p) = \frac{1}{2\pi\hbar}\int_{-\infty}^{\infty} e^{ipy/\hbar}\psi^*\left(x+\frac{y}{2}\right)\psi\left(x-\frac{y}{2}\right)dy \ , \tag{20}$$

which expresses quantum states as functions of position and momentum.

The Wigner distribution $W(x,p)$ is real, but it can be negative. Hence, it cannot be regarded as a probability distribution. Nevertheless, when integrated over one of the two variables, it yields the probability distribution for the other (for example, $\int W(x,p)dp = |\psi(x)|^2$. For a minimum uncertainty wave packet, $\psi(x) = \pi^{-1/4}\delta^{-1/2}\exp\{-(x-x_0)^2/2\delta^2 + ip_0x/\hbar\}$, the Wigner distribution is a Gaussian in both $x$ and $p$:

$$W(x,p) = \frac{1}{\pi\hbar}\exp\left\{-\frac{(x-x_0)^2}{\delta^2} - \frac{(p-p_0)^2\delta^2}{\hbar^2}\right\} \ . \tag{21}$$

It describes a system that is localized in both $x$ and $p$. Nothing else that Hilbert space has to offer is closer to approximating a point in classical phase space. The Wigner distribution is easily generalized to the case of a general density matrix $\rho(x,x')$:

$$W(x,p) = \frac{1}{2\pi\hbar}\int_{-\infty}^{-\infty} e^{ipy/\hbar}\rho\left(x-\frac{y}{2}, x+\frac{y}{2}\right)dy \ , \tag{22}$$

where $\rho(x,x')$ is, for example, the reduced density matrix of the particle discussed before.

The phase-space nature of the Wigner transform suggests a strategy for exhibiting classical behavior: Whenever $W(x,p)$ represents a mixture of localized wave packets—as in Equation (21)—it can be regarded as a classical probability distribution in the phase space. However, when the underlying state is truly quantum, as is the superposition in Figure 2, the corresponding Wigner distribution function will have alternating sign—see Figure 4(a). This property alone will make it impossible to regard the function as a probability distribution in phase space. The Wigner function in Figure 4(a) is

$$W(x,p) \sim \frac{\left(W^+ + W^-\right)}{2} + \frac{1}{\pi\hbar}\exp\left\{-\frac{p^2\delta^2}{\hbar^2} - \frac{x^2}{\delta^2}\right\}\cdot\cos\left(\frac{\Delta x}{\hbar}p\right) \ , \tag{23}$$

where the Gaussians $W^+$ and $W^-$ are Wigner transforms of the Gaussian wave packets $\chi^+$ and $\chi^-$. If the underlying state had been a mixture of $\chi^+$ and $\chi^-$ rather than a superposition, the Wigner function would have been described by the same two Gaussians $W^+$ and $W^-$, but the oscillating term would have been absent.

**Figure 4. Wigner Distributions and Their Decoherence for Coherent Superpositions**
**(a) The Wigner distribution $W(x, p)$ is plotted as a function of $x$ and $p$ for the cat state of Figure 2. Note the two separate positive peaks as well as the oscillating interference term in between them. This distribution cannot be regarded as a classical probability distribution in phase space because it has negative contributions. (b–e) Decoherence produces diffusion in the direction of the momentum. As a result, the negative and positive ripples of the interference term in $W(x, p)$ diffuse into each other and cancel out. This process is almost instantaneous for open macroscopic systems. In the appropriate limit, the Wigner function has a classical structure in phase space and evolves in accord with the equations of classical dynamics. (a′–e′) The analogous initial Wigner distribution and its evolution for a superposition of momenta are shown. The interference terms disappear more slowly on a time scale dictated by the dynamics of the system: Decoherence is caused by the environment coupling to (that is, monitoring) the position of the system—see Equation(16). So, for a superposition of momenta, it will start only after different velocities move the two peaks into different locations.**

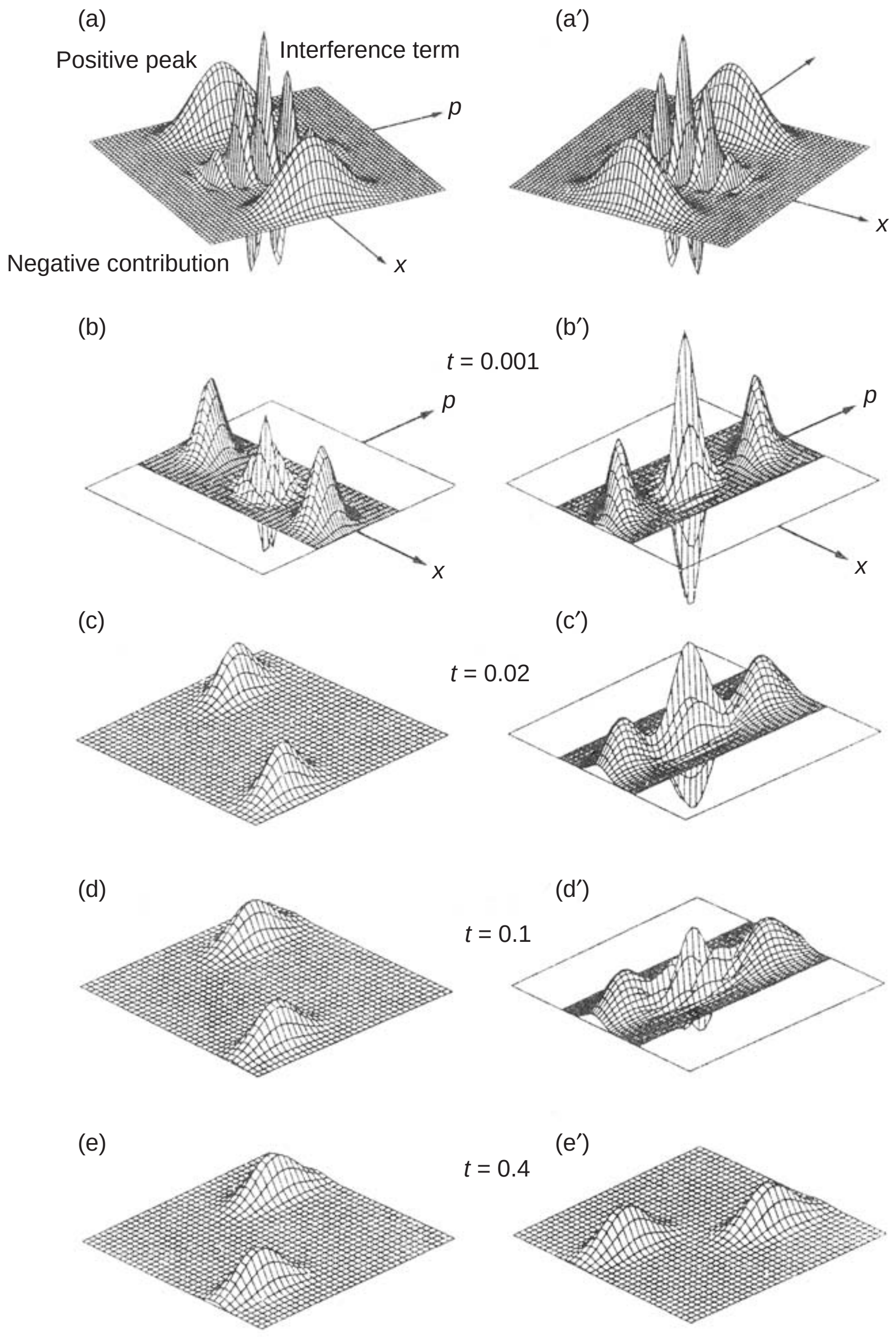

The equation of motion for $W(x, p)$ of a particle coupled to an environment can be obtained from Equation (17) for $\rho(x, x')$:

$$\frac{\partial W}{\partial t} = \underbrace{-\frac{p}{m}\frac{\partial}{\partial x}W + \frac{\partial V}{\partial x}\frac{\partial}{\partial p}W}_{Liouville\ \ Equation} + \underbrace{2\gamma\frac{\partial}{\partial p}pW}_{Friction} + \underbrace{D\frac{\partial^2 W}{\partial p^2}}_{Decoherence} , \tag{24}$$

where $V$ is the renormalized potential and $D = 2m\gamma k_B T = \eta k_B T$. The three terms of this equation correspond to the three terms of Equation (17).

The first term is easily identified as a classical Poisson bracket $\{H, W\}$. That is,

### The Predictability Sieve

Since 1991, understanding the emergence of preferred pointer states during the process of decoherence has advanced a great deal. Perhaps the most important advance is the predictability sieve (Zurek 1993, Zurek et al. 1993), a more general definition of pointer states that obtains even when the interaction with the environment does not dominate over the self-Hamiltonian of the system. The predictability sieve sifts through the Hilbert space of a system interacting with its environment and selects states that are most predictable. Motivation for the predictability sieve comes from the observation that classical states exist or evolve predictably. Therefore, selecting quantum states that retain predictability in spite of the coupling to the environment is the obvious strategy in search of classicality. To implement the predictability sieve, we imagine a (continuously infinite) list of all the pure states $\{|\psi\rangle\}$ in the Hilbert space of the system in question. Each of them would evolve, after a time $t$, into a density matrix $\rho_{|\psi\rangle}(t)$. If the system were isolated, all the density matrices would have the form $\rho_{|\psi\rangle}(t) = |\psi(t)\rangle\langle\psi(t)|$ of projection operators, where $|\psi(t)\rangle$ is the appropriate solution of the Schrödinger equation. But when the system is coupled to the environment (that is, the system is "open"), $\rho_{|\psi\rangle}(t)$ is truly mixed and has a nonzero von Neumann entropy. Thus, one can compute $\mathcal{H}(\rho_{|\psi\rangle}(t)) = -Tr\rho_{|\psi\rangle}(t)\log\rho_{|\psi\rangle}(t)$, thereby defining a functional on the Hilbert space $\mathcal{H}_\mathcal{S}$ of the system, $|\psi\rangle \rightarrow \mathcal{H}(|\psi\rangle, t)$.

An obvious way to look for predictable, effectively classical states is to seek a subset of all $\{|\psi\rangle\}$ that minimize $\mathcal{H}(|\psi\rangle, t)$ after a certain, sufficiently long time $t$. When such preferred pointer states exist, are well defined (that is, the minimum of the entropy $\mathcal{H}(|\psi\rangle, t)$ differs significantly for pointer states from the average value), and are reasonably stable (that is, after the initial decoherence time, the set of preferred states is reasonably insensitive to the precise value of $t$), one can consider them as good candidates for the classical domain. Figure A illustrates an implementation of the predictability sieve strategy using a different, simpler measure of predictability—purity ($\mathrm{Tr}\rho^2$)—rather than the von Neumann entropy, which is much more difficult to compute.

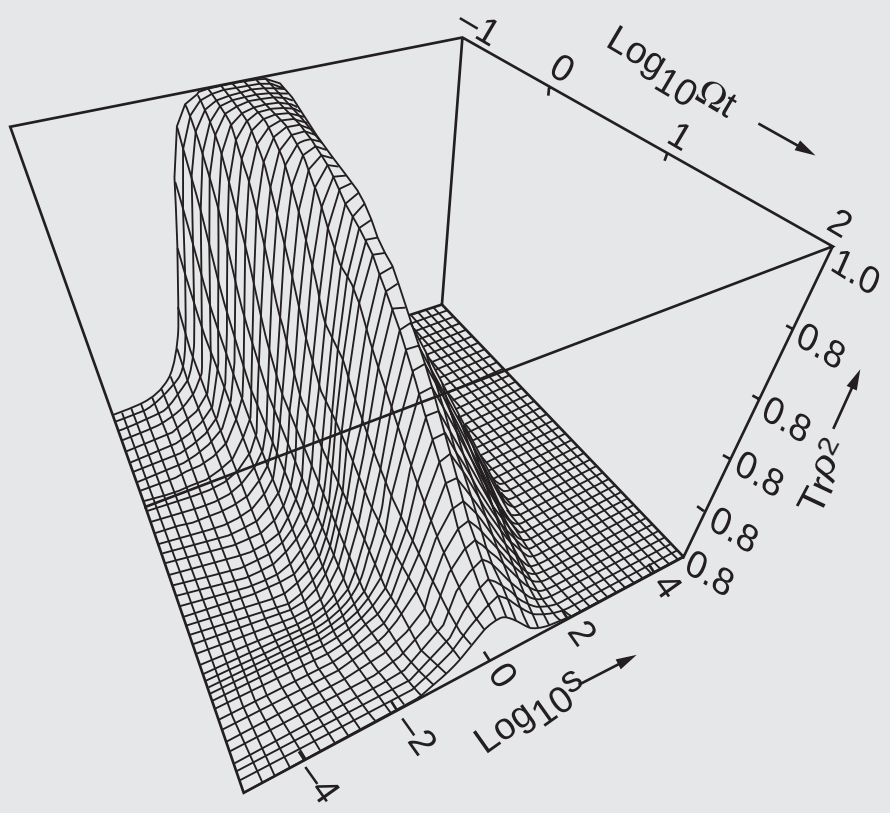

**Figure A. The Predictability Sieve for the Underdamped Harmonic Oscillator**
**One measure of predictability is the so-called purity $\mathrm{Tr}\rho^2$, which is plotted as a function of time for mixtures of minimum uncertainty wave packets in an underdamped harmonic oscillator with $\gamma/\omega = 10^{-4}$. The wave packets start with different squeeze parameters $s$. $\mathrm{Tr}\rho^2$ serves as a measure of the purity of the reduced density matrix $\rho$. The predictability sieve favors coherent states ($s$ = 1), which have the shape of a ground state, that is, the same spread in position and momentum when measured in units natural for the harmonic oscillator. Because they are the most predictable (more than the energy eigenstates), they are expected to play the crucial role of the pointer basis in the transition from quantum to classical.**

if $\rho(x, p)$ is a familiar classical probability density in phase space, then it evolves according to:

$$\frac{\partial w}{\partial t} = -\frac{\partial w}{\partial x}\frac{\partial H}{\partial p} + \frac{\partial w}{\partial p}\frac{\partial H}{\partial x} = \{H, w\} = Lw \qquad (25)$$

where $L$ stands for the Liouville operator. Thus, classical dynamics in its Liouville form follows from quantum dynamics at least for the harmonic oscillator case, which is described rigorously by Equations (17) and (24). (For more general $V(x)$, the Poisson bracket would have to be supplemented by quantum corrections of order $\hbar$.) The second term of Equation (24) represents friction. The last term results in the diffusion of $W(x, p)$ in momentum at the rate given by $D$.

## Quantum Chaos and Phase-Space Aspects of the Quantum-Classical Correspondence

Classical mechanics "happens" in phase space. It is therefore critically important to show that quantum theory can—in the presence of decoherence—reproduce the basic structure of classical phase space and that it can emulate classical dynamics. The argument put forward in my original paper (1991) has since been amply supported by several related developments.

The crucial idealization that plays a key role in classical physics is a "point." Because of Heisenberg's principle, $\Delta x\, \Delta p \geq \hbar/2$, quantum theory does not admit states with simultaneously vanishing $\Delta x$ and $\Delta p$. However, as the study of the predictability sieve has demonstrated, in many situations relevant to the classical limit of quantum dynamics, one can expect decoherence to select pointer states that are localized in both $\Delta x$ and $\Delta p$, that is, approximate minimum uncertainty wave packets. In effect, these wave packets are a quantum version of points, which appear naturally in the underdamped harmonic oscillator coupled weakly to the environment (Zurek et al. 1993, Gallis 1996). These results are also relevant to the transition from quantum to classical in the context of field theory with the added twist that the kinds of states selected will typically differ for bosonic and fermionic fields (Anglin and Zurek 1996) because bosons and fermions tend to couple differently to their environments. Finally, under suitable circumstances, einselection can even single out energy eigenstates of the self-Hamiltonian of the system, thus justifying in part the perception of "quantum jumps" (Paz and Zurek 1999).

Quantum chaos provides an intriguing arena for the discussion of the quantum-classical correspondence. To begin with, classical and quantum evolutions from the same initial conditions of a system lead to very different phase-space "portraits." The quantum phase-space portrait will depend on the particular representation used, but there are good reasons to favor the Wigner distribution. Studies that use the Wigner distribution indicate that, at the moment when the quantum-classical correspondence is lost in chaotic dynamics, even the averages computed using properties of the classical and quantum states begin to differ (Karkuszewski et al. 2002).

Decoherence appears to be very effective in restoring correspondence. This point, originally demonstrated almost a decade ago (Zurek and Paz 1994, 1995) has since been amply corroborated by numerical evidence (Habib et al. 1998). Basically, decoherence eradicates the small-scale interference accompanying the rapid development of large-scale coherence in quantum versions of classically chaotic systems (refer to Figure A). This outcome was expected. In order for the quantum to classical correspondence to hold, the coherence length $\ell_C$ of the quantum state must satisfy the following inequality: $\ell_C = \hbar/(2D\lambda)^{1/2} << \chi$, where $\lambda$ is the Lyapunov exponent, $D$ is the usual coefficient describing the rate of decoherence, and $\chi$ is the scale on which the potential $V(x)$ is significantly nonlinear:

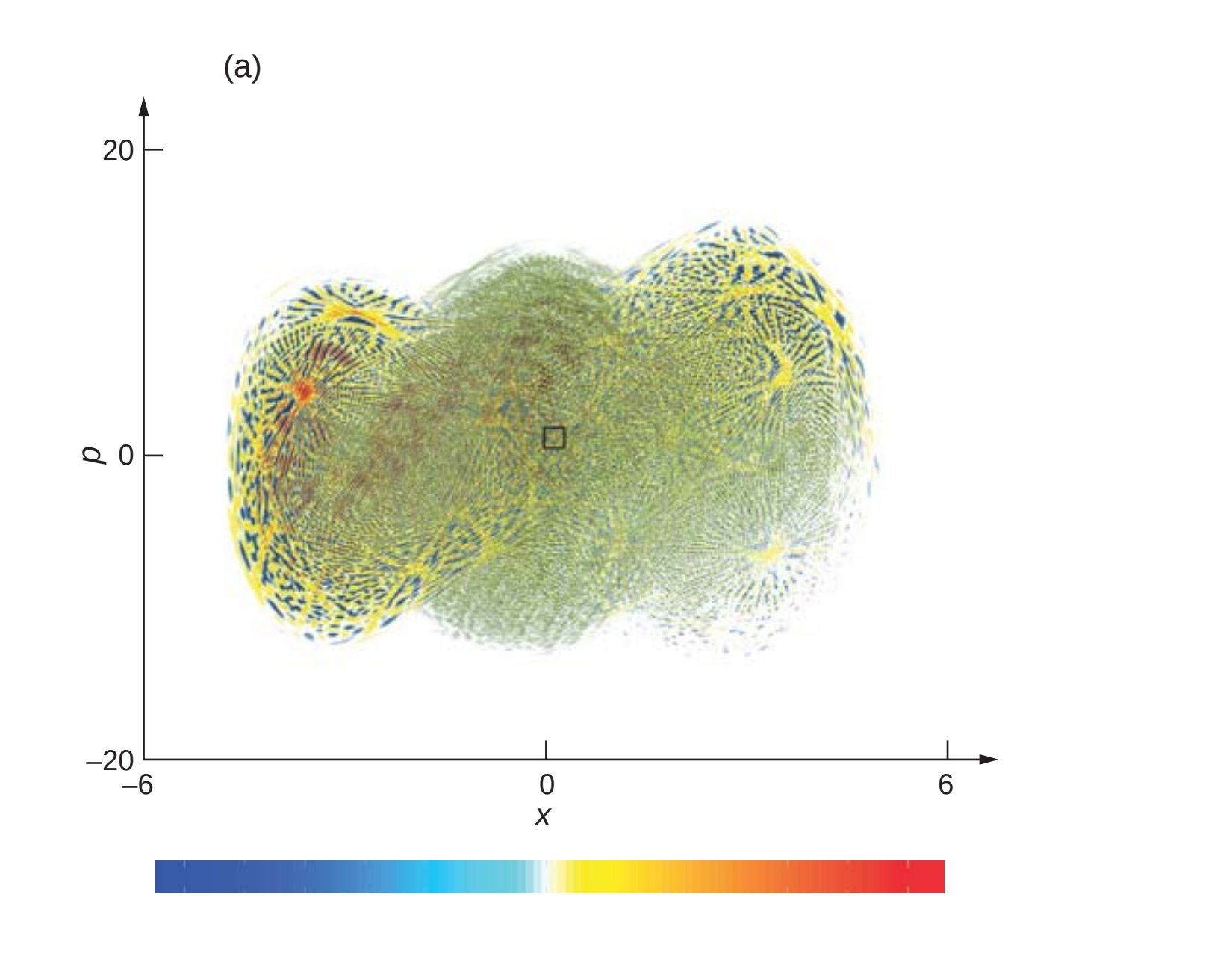

$$\chi \cong \sqrt{\frac{V'}{V'''}} \quad .$$

When a quantum state is localized on scales small compared to $\chi$ (which is the import of the inequality above), its phase space evolution is effectively classical, but because of chaos and decoherence, it becomes irreversible and unpredictable. Nevertheless—as argued in this volume by Tanmoy Bhattacharya, Salman Habib, and Kurt Jacobs (on page XX)—one can even recover more or less classical trajectories by modeling a continuous measurement. However, this is an extra ingredient not in the spirit of the decoherence approach as it invokes the measurement process without explaining it.

A surprising corollary of this line of argument is the realization (Zurek and Paz 1994) that the dynamical second law—entropy production at the scale set by the dynamics of the system and more or less independent of the strength of the coupling to the environment—is a natural and, indeed, an inevitable consequence of decoherence. This point has been since confirmed in numerical studies (Miller and Sarkar 1999, Pattanayak 1999, Monteoliva and Paz 2000).

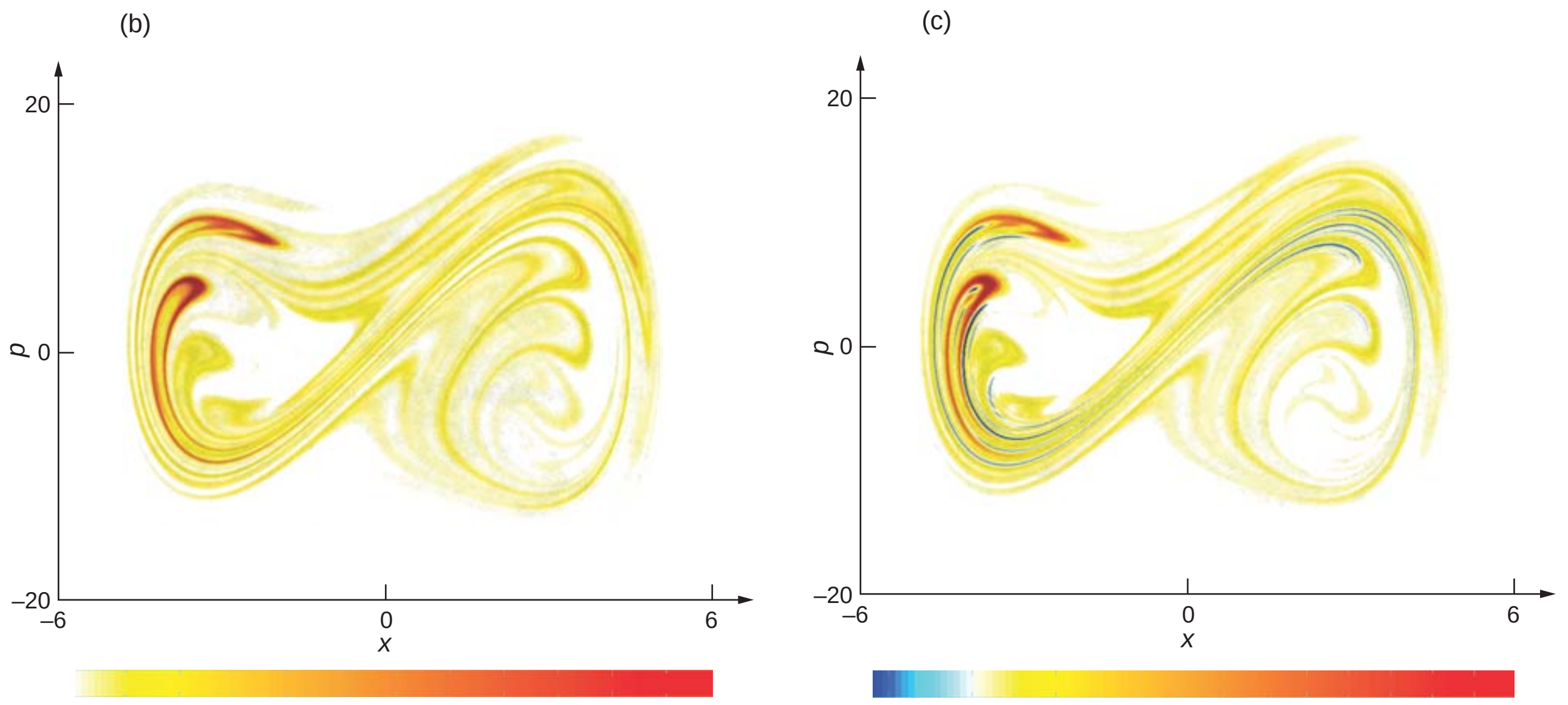

**Figure A. Decoherence in a Chaotic Driven Double-Well System**
**This numerical study (Habib et al. 1998) of a chaotic driven double-well system described by the Hamiltonian $H = p^2/2m - Ax^2 + Bx^4 + Fx\cos(\omega t)$ with $m = 1$, $A = 10$, $B = 0.5$, $F = 10$, and $\omega = 6.07$ illustrates the effectiveness of decoherence in the transition from quantum to classical. These parameters result in a chaotic classical system with a Lyapunov exponent $\lambda \cong 0.5$. The three snapshots taken after 8 periods of the driving force illustrate phase space distributions in (a) the quantum case, (b) the classical case, and (c) the quantum case but with decoherence ($D = 0.025$). The initial condition was always the same Gaussian, and in the quantum cases, the state was pure. Interference fringes are clearly visible in (a), which bears only a vague resemblance to the classical distribution in (b). By contrast, (c) shows that even modest decoherence helps restore the quantum-classical correspondence. In this example the coherence length $\ell_C$ is smaller than the typical nonlinearity scale, so the system is on the border between quantum and classical. Indeed, traces of quantum interference are still visible in (c) as blue "troughs," or regions where the Wigner function is still slightly negative. The change in color from red to blue shown in the legends for (a) and (c) corresponds to a change from positive peaks to negative troughs. In the ab initio classical case (b), there are no negative troughs.**

Other surprising consequences of the study of Wigner functions in the quantum-chaotic context is the realization that they develop phase space structure on the scale associated with the sub-Planck action $a = \hbar^2/A << \hbar$, where $A$ is the classical action of the system, and that this Planck action is physically significant (Zurek 2001b). This can be seen in Figure A part (a), where a small black square with the area of $\hbar$ is clearly larger than the smallest "ripples" in the image.

This point was to some extent anticipated by the plots of the Wigner functions of Schrödinger cats (see Figures 4a and 4a′ in this article)a version of which appeared in the 1991 *Physics Today* version of this paper—the interference term of the Wigner function has a sub-Planck structure.

A lot has happened in establishing phase-space aspects of quantum-classical correspondence, but a lot more remains to be done. (A more thorough summary of the past accomplishments and remaining goals can be found in Zurek 2001b).

Classical equations of motion are a necessary but insufficient ingredient of the classical limit: We must also obtain the correct structure of the classical phase space by barring all but the probability distributions of well-localized wave packets. The last term in Equation (24) has precisely this effect on nonclassical $W(x,p)$. For example, the Wigner function for the superposition of spatially localized wave packets—Figure 4(a)—has a sinusoidal modulation in the momentum coordinate produced by the oscillating term $\cos((\Delta x/\hbar)p)$. This term, however, is an eigenfunction of the diffusion operator $\partial^2/\partial p^2$ in the last term of Equation (24). As a result, the modulation is washed out by diffusion at a rate

$$\tau_D^{-1} = -\frac{\dot{W}}{W} = \frac{\left(D\dfrac{\partial^2}{\partial p^2}W\right)}{W} = \frac{2m\gamma k_B T\left(\Delta x\right)^2}{\hbar^2} .$$

Negative valleys of $W(x,p)$ fill in on a time scale of order $\tau_D$, and the distribution retains just two peaks, which now correspond to two classical alternatives—see Figures 4(a) to 4(e). The Wigner function for a superposition of momenta, shown in Figure 4(a′), also decoheres as the dynamics causes the resulting difference in velocities to damp out the oscillations in position and again yield two classical alternatives—see Figures 4(b′) to 4(e′).

The ratio of the decoherence and relaxation time scales depends on $\hbar^2/m$—see Equation (19). Therefore, when $m$ is large and $\hbar$ small, $\tau_D$ can be nearly zero—decoherence can be nearly instantaneous—while, at the same time, the motion of small patches (which correspond to the probability distribution in classical phase space) in the smooth potential becomes reversible. This idealization is responsible for our confidence in classical mechanics, and, more generally, for many aspects of our belief in classical reality.

The discussion above demonstrates that decoherence and the transition from quantum to classical (usually regarded as esoteric) is an inevitable consequence of the immersion of a system in an environment. True, our considerations were based on a fairly specific model—a particle in a heat bath of harmonic oscillators. However, this is often a reasonable approximate model for many more complicated systems. Moreover, our key conclusions—such as the relation between the decoherence and relaxation time scales in Equation (19)—do not depend on any specific features of the model. Thus, one can hope that the viscosity and the resulting relaxation always imply decoherence and that the transition from quantum to classical can be always expected to take place on a time scale of the order of the above estimates.

## Quantum Theory of Classical Reality

Classical reality can be defined purely in terms of classical states obeying classical laws. In the past few sections, we have seen how this reality emerges from the substrate of quantum physics: Open quantum systems are forced into states described by localized wave packets. They obey classical equations of motion, although with damping terms and fluctuations that have a quantum origin. What else is there to explain?

Controversies regarding the interpretation of quantum physics originate in the clash between the predictions of the Schrödinger equation and our perceptions. I will therefore conclude this paper by revisiting the source of the problem—our awareness of definite outcomes. If these mental processes were essentially unphysical, there would be no hope of formulating and addressing the ultimate question—why do we perceive just one of the quantum alternatives?—within the context of physics. Indeed, one might be tempted to follow Eugene Wigner (1961) and give consciousness the last word in collapsing the state vector. I shall assume the opposite. That is, I shall examine the idea that the higher mental processes all correspond to well-defined, but at present, poorly understood information-processing functions that are being carried out by physical systems, our brains.

Described in this manner, awareness becomes susceptible to physical analysis. In particular, the process of decoherence we have described above is bound to affect the states of the brain: Relevant observables of individual neurons, including chemical concentrations and electrical potentials, are macroscopic. They obey classical, dissipative equations of motion. Thus, any quantum superposition of the states of neurons will be destroyed far too quickly for us to become conscious of the quantum "goings on." Decoherence, or more to the point, environment-induced superselection, applies to our own "state of mind."

One might still ask why the preferred basis of neurons becomes correlated with the classical observables in the familiar universe. It would be, after all, so much easier to believe in quantum physics if we could train our senses to perceive nonclassical superpositions. One obvious reason is that the selection of the available interaction Hamiltonians is limited and constrains the choice of detectable observables. There is, however, another reason for this focus on the classical that must have played a decisive role: Our senses did not evolve for the purpose of verifying quantum mechanics. Rather, they have developed in the process in which survival of the fittest played a central role. There is no evolutionary reason for perception when nothing can be gained from prediction. And, as the predictability sieve illustrates, only quantum states that are robust in spite of decoherence, and hence, effectively classical, have predictable consequences. Indeed, classical reality can be regarded as nearly synonymous with predictability.

There is little doubt that the process of decoherence sketched in this paper is an important element of the big picture central to understanding the transition from quantum to classical. Decoherence destroys superpositions. The environment induces, in effect, a superselection rule that prevents certain superpositions from being observed. Only states that survive this process can become classical.

There is even less doubt that this rough outline will be further extended. Much work needs to be done both on technical issues (such as studying more realistic models that could lead to additional experiments) and on problems that require new conceptual input (such as defining what constitutes a "system" or answering the question of how an observer fits into the big picture).

Decoherence is of use within the framework of either of the two interpretations: It can supply a definition of the branches in Everett's Many Worlds Interpretation, but it can also delineate the border that is so central to Bohr's point of view. And if there is one lesson to be learned from what we already know about such matters, it is that information and its transfer play a key role in the quantum universe.

The natural sciences were built on a tacit assumption: Information about the universe can be acquired without changing its state. The ideal of "hard science" was to be objective and provide a description of reality. Information was regarded as unphysical, ethereal, a mere record of the tangible, material universe, an inconsequential reflection, existing beyond and essentially decoupled from the domain governed by the laws of physics. This view is no longer tenable (Landauer 1991). Quantum theory has put an end to this Laplacean dream about a mechanical universe. Observers of quantum phenomena can no longer be just passive spectators. Quantum laws make it impossible to gain information without changing the state of the measured object. The dividing line between what is and what is known to be has been blurred forever. While abolishing this boundary, quantum theory has simultaneously deprived the "conscious observer" of a monopoly on acquiring and storing information: Any correlation is a registration, any quantum state is a record of some other quantum state. When correlations are robust enough, or the record is sufficiently indelible, familiar classical "objective reality" emerges from the quantum substrate. Moreover, even a minute interaction with the environment, practically inevitable for any macroscopic object, will establish such a correlation: The environment will, in effect, measure the state of the object, and this suffices to destroy quantum coherence. The resulting decoherence plays, therefore, a vital role in facilitating the transition from quantum to classical.

## The Existential Interpretation

The quantum theory of classical reality has developed significantly since 1991. These advances are now collectively known as the existential interpretation (Zurek 2001a). The basic difference between quantum and classical states is that the objective existence of the latter can be taken for granted. That is, a system's classical state can be simply "found out" by an observer originally ignorant of any of its characteristics. By contrast, quantum states are hopelessly "malleable"—it is impossible in principle for an observer to find out an unknown quantum state without perturbing it. The only exception to this rule occurs when an observer knows beforehand that the unknown state is one of the eigenstates of some definite observable. Then and only then can a nondemolition measurement (Caves et al. 1980) of that observable be devised such that another observer who knew the original state would not notice any perturbations when making a confirmatory measurement.

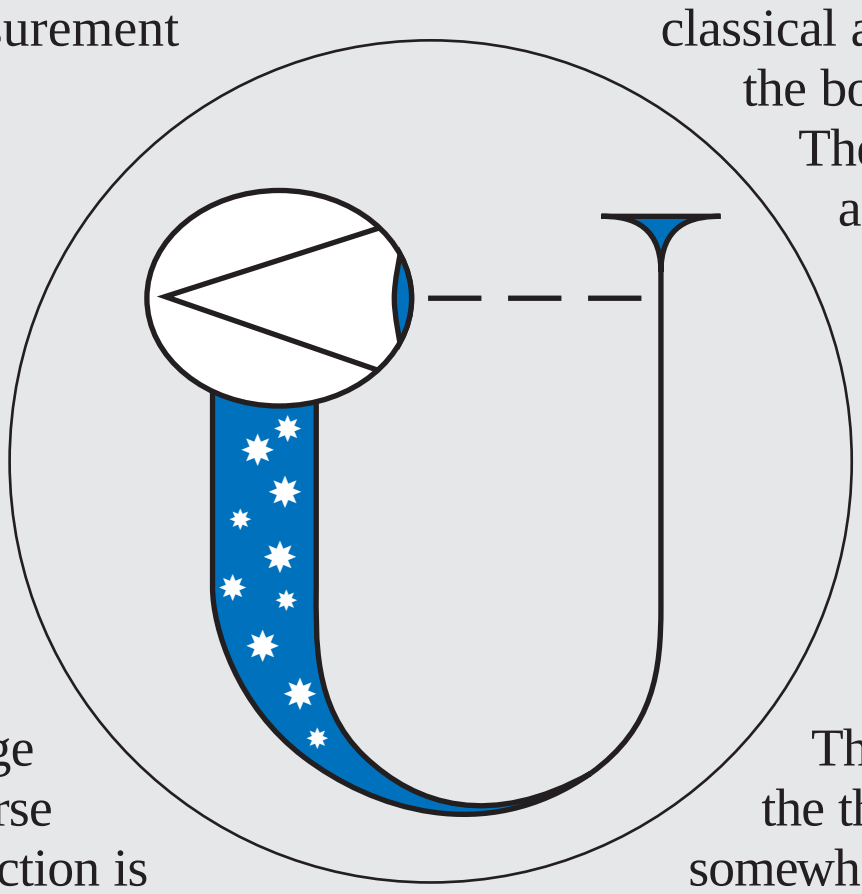

If the unknown state cannot be found out—as is indeed the case for isolated quantum systems—then one can make a persuasive case that such states are subjective, and that quantum state vectors are merely records of the observer's knowledge about the state of a fragment of the Universe (Fuchs and Peres 2000). However, einselection is capable of converting such malleable and "unreal" quantum states into solid elements of reality. Several ways to argue this point have been developed since the early discussions (Zurek 1993, 1998, 2001a). In effect, all of them rely on einselection, the emergence of the preferred set of pointer states. Thus, observers aware of the structure of the Hamiltonians (which are "objective," can be found out without "collateral damage", and in the real world, are known well enough in advance) can also divine the sets of preferred pointer states (if they exist) and thus discover the preexisting state of the system.

One way to understand this environment-induced objective existence is to recognize that observers—especially human observers—never measure anything directly. Instead, most of our data about the Universe is acquired when information about the systems of interest is intercepted and spread throughout the environment. The environment preferentially records the information about the pointer states, and hence, only information about the pointer states is readily available. This argument can be made more rigorous in simple models, whose redundancy can be more carefully quantified (Zurek 2000, 2001a).

This is an area of ongoing research. Acquisition of information about the systems from fragments of the environment leads to the so-called conditional quantum dynamics, a subject related to quantum trajectories (Carmichael 1993). In particular one can show that the predictability sieve also works in this setting (Dalvit et al. 2001).

The overarching open question of the interpretation of quantum physics—the "meaning of the wave function"—appears to be in part answered by these recent developments. Two alternatives are usually listed as the only conceivable answers. The possibility that the state vector is purely epistemological (that is, solely a record of the observer's knowledge) is often associated with the Copenhagen Interpretation (Bohr 1928). The trouble with this view is that there is no unified description of the Universe as a whole: The classical domain of the Universe is a necessary prerequisite, so both classical and quantum theory are necessary and the border between them is, at best, ill-defined. The alternative is to regard the state vector as an ontological entity—as a solid description of the state of the Universe akin to the classical states. But in this case (favored by the supporters of Everett's Many Worlds Interpretation), everything consistent with the universal state vector needs to be regarded as equally "real."

The view that seems to be emerging from the theory of decoherence is in some sense somewhere in between these two extremes. Quantum state vectors can be real, but only when the superposition principle—a cornerstone of quantum behavior—is "turned off" by einselection. Yet einselection is caused by the transfer information about selected observables. Hence, the ontological features of the state vectors—objective existence of the einselected states—is acquired through the epistemological "information transfer."

Obviously, more remains to be done. Equally obviously, however, decoherence and einselection are here to stay. They constrain the possible solutions after the quantum–classical transition in a manner suggestive of a still more radical view of the ultimate interpretation of quantum theory in which information seems destined to play a central role. Further speculative discussion of this point is beyond the scope of the present paper, but it will be certainly brought to the fore by (paradoxically) perhaps the most promising applications of quantum physics to information processing. Indeed, quantum computing inevitably poses questions that probe the very core of the distinction between quantum and classical. This development is an example of the unpredictability and serendipity of the process of scientific discovery: Questions originally asked for the most impractical of reasons—questions about the EPR paradox, the quantum-to-classical transition, the role of information, and the interpretation of the quantum state vector—have become relevant to practical applications such as quantum cryptography and quantum computation.■

## Acknowledgments

I would like to thank John Archibald Wheeler for many inspiring and enjoyable discussions on "the quantum" and Juan Pablo Paz for the pleasure of a long-standing collaboration on the subject.

## Further Reading

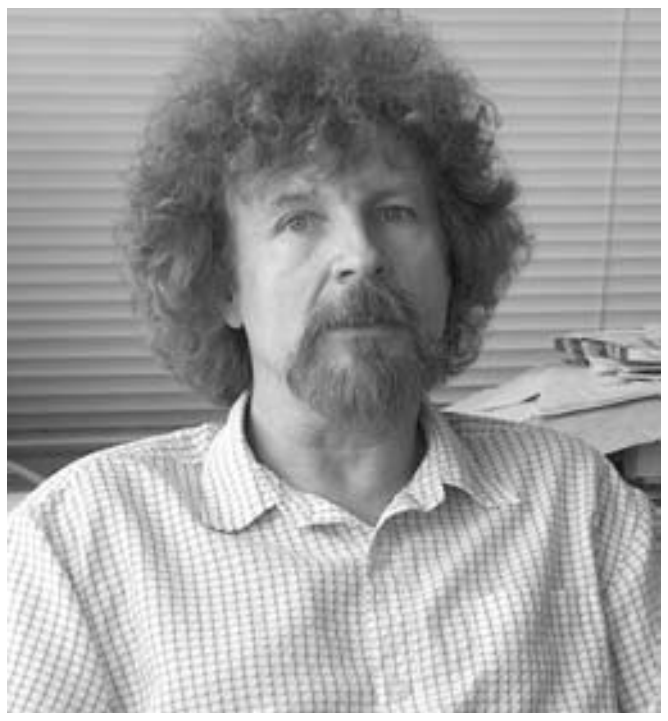

**Wojciech Hubert Zurek** was educated in Kraków, in his native Poland (M. Sc., 1974) and in Austin, Texas (Ph. D. in physics, 1979). He was a Richard Chace Tolman Fellow at the California Institute of Technology and a J. Robert Oppenheimer Fellow at the Los Alamos National Laboratory. In 1996, Wojciech was selected as a Los Alamos National Laboratory Fellow. He is a Foreign Associate of the Cosmology Program of the Canadian Institute of Advanced Research and the founder of the Complexity, Entropy, and Physics of Information Network of the Santa Fe Institute. His research interests include decoherence, physics of quantum and classical information, foundations of statistical and quantum physics and astrophysics.